\documentclass{article}

\usepackage{PRIMEarxiv}

\usepackage[utf8]{inputenc} 
\usepackage[T1]{fontenc}    

\usepackage[
    colorlinks=true,
    linkcolor=black,
    urlcolor=blue,
    citecolor=black
]{hyperref}
\usepackage{xcolor}
\usepackage{url}            
\usepackage{booktabs}       
\usepackage{amsfonts}       
\usepackage{nicefrac}       
\usepackage{microtype}      
\usepackage{amsmath}
\usepackage{lipsum}
\usepackage{fancyhdr}       
\usepackage{graphicx}       
\usepackage{fancyhdr}
\usepackage{datetime}
\usepackage[useregional]{datetime2}
\usepackage[utf8]{inputenc}
\usepackage[T1]{fontenc}
\usepackage{pgfplots}
\usepackage{float}
\usepackage{fontawesome5}
\usepackage{enumitem}
\setlist[itemize]{
    itemsep=3pt,
    topsep=4pt,
    parsep=0pt,
    partopsep=0pt
}
\usepackage{amssymb}
\pgfplotsset{compat=1.18}
\graphicspath{{media/}}     

\title{Tadabur: A Large-Scale Quran Dataset
\footnotetext{\textit{\underline{Citation}}: 
\textbf{Authors. Title. Pages.... DOI:000000/11111.}}
}

\author{
  Faisal Alherran \\
  Riyadh, Saudi Arabia\\
  alherranfaisal@gmail.com
}

\begin{document}
\thispagestyle{fancy}

\vspace*{0.4cm}

\begin{center}
    {\LARGE\bfseries Tadabur: A Large-Scale Quran Audio Dataset}\\[0.5cm]
    {\large Faisal Alherran}\\
    Riyadh, Saudi Arabia\\
    \texttt{alherranfaisal@gmail.com}
\end{center}

\vspace{0.1cm}



\begin{center}
\small
\href{https://github.com/fherran/tadabur}{%
  \textcolor[HTML]{1c1f1e}{\faGithub}~Github}
\quad\textcolor{gray}{|}  \quad
\href{https://huggingface.co/datasets/FaisaI/tadabur}{%
  \raisebox{-0.2\height}{\includegraphics[height=1.1em]{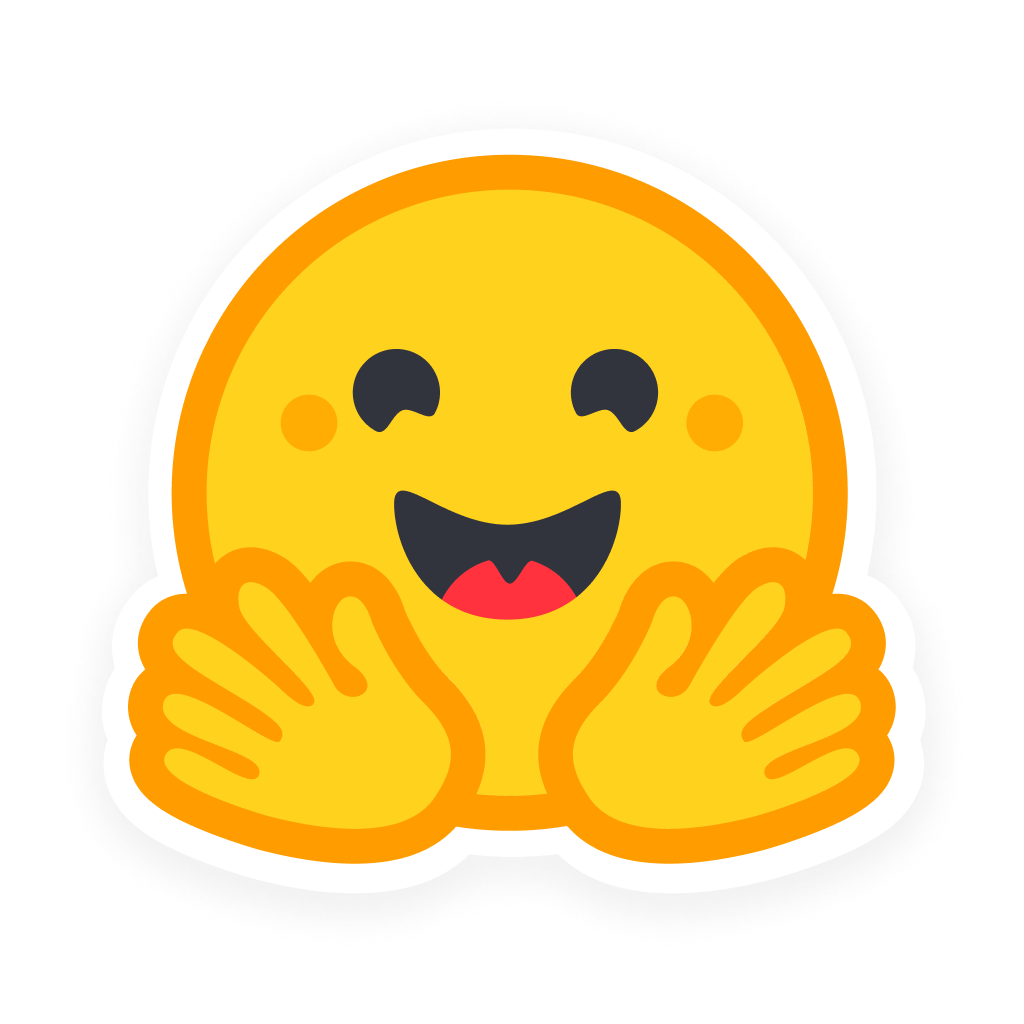}}~Huggingface}
\quad\textcolor{gray}{|}  \quad
\href{https://fherran.github.io/tadabur}{%
  \textcolor[HTML]{a27b5c}{\faGlobe}~Tadabur Page}
\end{center}
\vspace{0.5cm}

\begin{abstract}
Despite growing interest in Quranic data research, existing Quran datasets remain limited in both scale and diversity. To address this gap, we present \textbf{Tadabur}, a large-scale Quran audio dataset. Tadabur comprises more than \textbf{1400+ hours} of recitation audio from over \textbf{600} distinct reciters, providing substantial variation in recitation styles, vocal characteristics, and recording conditions. This diversity makes \textbf{Tadabur} a comprehensive and representative resource for Quranic speech research and analysis. By significantly expanding both the total duration and variability of available Quran data, Tadabur aims to support future research and facilitate the development of standardized Quranic speech benchmarks.
\end{abstract}

\section{Introduction}
Audio understanding plays a central role in modern machine learning, yet Qur'anic audio---despite its global significance and unique acoustic properties---remains underrepresented in research. Existing Qur'anic speech datasets are limited in scale, reciter diversity, audio quality, and annotation depth, restricting progress in tasks such as Qur'anic automatic speech recognition (ASR), tajwīd-aware modeling, reciter identification, and prosodic analysis. As a result, current systems often fail to capture the rich stylistic variation, strict phonological rules, and melodic structures that characterize Qur'anic recitation.

To address these limitations, we introduce \textbf{Tadabur}, a large and diverse Qur'anic audio dataset. Tadabur comprises more than \textbf{1400+ hours} of audio from over \textbf{600} distinct reciters, with complete coverage of \textbf{113} surahs excluding Al-Fatiha and thousands of Qur'anic verses. The dataset spans a wide range of recitation styles (e.g., \textit{murattal}, \textit{mujawwad}), speaking rates, recording conditions, and acoustic qualities, and is accompanied by automatically derived metadata and precise temporal annotations.

These characteristics make Tadabur one of the most comprehensive and representative resources available for Qur'anic speech research. The dataset enables advancements in ASR and speech modeling, large-scale speaker and style analysis, prosody and tajwīd research, and data-centric studies on robustness and transfer learning. By establishing a standardized and richly varied benchmark, Tadabur provides a strong foundation for future work in domain-adapted speech technologies.

In summary, this work makes the following contributions:
\begin{itemize}
    \item We introduce \textbf{Tadabur}, a large-scale Qur'anic speech dataset comprising more than \textbf{1400+ hours} of audio from over \textbf{600} reciters.
    \item We propose an automated pipeline for large-scale Qur'an data curation that combines LLM-based metadata extraction, Whisper/WhisperX-based alignment, and ASR-driven content filtering to obtain high-quality, time-aligned annotations.
    \item We provide machine-readable word-level alignments and structured metadata for each verse-level audio file in a consistent JSON schema.
\end{itemize}

\section{Related Work}
\label{sec:related_work}
\subsection{Quran Dataset}
Several Quranic audio datasets have been introduced in recent years to support research in automatic speech recognition (ASR), pronunciation assessment, and computer-assisted Quran recitation. However, despite these efforts, most existing datasets remain limited in overall scale, reciter and speaker diversity, recording condition variability, and the richness of linguistic and phonetic annotations. The most notable publicly available datasets can be summarized as follows:

\begin{itemize}
    \item \textbf{Quran Recitations for Audio Classification \cite{alrajeh2023quranrecitations}:} 
    Sourced publicly from Kaggle, this dataset was originally curated for the task of reciter identification rather than speech recognition. It consists of a modest collection of 6,689 audio files attributed to 12 distinct reciters. While the audio samples capture different acoustic environments, the dataset lacks the fundamental linguistic annotations required for ASR training. Specifically, it does not include textual transcriptions or time-aligned metadata. Consequently, its utility is strictly limited to audio classification tasks (e.g., speaker identification) and it is unsuitable for training acoustic models to recognize Quranic content.

    \item \textbf{Quran Speech to Text Dataset (SLR132) \cite{openslr132_quran}:} 
    As one of the standard benchmarks for Quranic ASR, the SLR132 corpus provides a structured collection of 226,129 audio-text pairs sourced from 30 renowned reciters. It aligns audio at the \textit{ayah} (verse) level with corresponding Arabic text, making it a viable baseline for developing verse-level recognition systems. However, its limitation lies in its coarse granularity; the lack of word-level or phoneme-level timestamping restricts its application in more advanced tasks, such as misalignment detection or fine-grained pronunciation scoring. Furthermore, the reliance on a small set of professional reciters limits the acoustic diversity necessary for training models that generalize well to non-professional or diverse speakers.

    \item \textbf{Buraaq (Quran Audio-Text Dataset) \cite{salman2025quran}:} 
    Hosted on Hugging Face, the Buraaq dataset comprises approximately 187,080 samples derived from 30 reciters. It distinguishes itself by including rich metadata, such as translations and surah information, which supports multi-task learning scenarios.
\end{itemize}

\subsection{Automatic Speech Recognition (ASR)}

Automatic Speech Recognition (ASR) has progressed rapidly over the past decade, driven by advances in deep learning, large-scale datasets, and end-to-end modeling frameworks. Early ASR systems were predominantly based on hybrid Hidden Markov Model–Gaussian Mixture Model (HMM–GMM) architectures, which relied heavily on handcrafted acoustic features and complex modular pipelines. The transition to deep neural networks (DNNs), including convolutional neural networks (CNNs), recurrent neural networks (RNNs), and later Long Short-Term Memory (LSTM) and Gated Recurrent Unit (GRU) architectures, significantly improved acoustic modeling and recognition accuracy.

The introduction of Connectionist Temporal Classification (CTC) enabled alignment-free sequence training, reducing reliance on frame-level annotations. This was followed by attention-based encoder–decoder architectures, which unified acoustic and language modeling into a single end-to-end framework. More recently, Transformer-based architectures have become the dominant paradigm due to their ability to model long-range dependencies and their scalability to massive datasets.

A major shift in contemporary ASR research is the adoption of self-supervised representation learning. Models such as wav2vec~2.0, HuBERT, and Whisper leverage large amounts of unlabeled speech to learn rich and transferable acoustic representations, which can then be fine-tuned with limited labeled data. These methods have achieved state-of-the-art performance across diverse languages, acoustic conditions, and speaking styles, highlighting the central role of large and diverse datasets in achieving robust ASR systems.

Despite these advances, several challenges remain, particularly for highly structured and stylistically rich speech domains. In the context of Qur’anic recitation, ASR systems must contend with prolonged phoneme durations, strict pronunciation rules (tajwīd), melodic articulation, speaker-dependent recitation styles, and significant acoustic variability across recording environments. Furthermore, most large-scale ASR benchmarks are based on conversational or read speech and do not adequately represent the linguistic and prosodic characteristics of Qur’anic audio.

The availability of a large and highly variant Qur’an speech dataset therefore plays a critical role in advancing ASR performance in this domain. Such datasets enable robust modeling of reciter diversity, recitation styles, and pronunciation variability, while also supporting the development of domain-adapted self-supervised and end-to-end ASR systems. This work is directly motivated by these challenges and aims to bridge the gap between general-purpose ASR models and the unique characteristics of Qur’anic speech.

\section{Dataset Overview}

\textbf{Tadabur} dataset is a large-scale Qur’anic speech corpus compiled from a diverse collection of well-known public Qur’an audio publishers. It is designed to capture wide variation across reciters, recitation styles, chapters, acoustic environments, and recording qualities. Unlike conventional speech corpora that primarily focus on conversational or read speech, Tadabur targets a highly structured and prosodically rich domain governed by strict phonological and rhythmic rules. The dataset construction pipeline follows a multi-stage, fully automated process consisting of data collection, metadata extraction, verse-level alignment, automated content cleaning, and final validation to ensure both linguistic and acoustic integrity.

\subsection{Data Collection}

Audio data were collected from publicly accessible Qur’an repositories and online publishing platforms that host large-scale recitation archives from numerous reciters. The collection strategy was designed to maximize diversity across several critical dimensions, including reciter identity, recitation style (e.g., murattal, mujawwad), recording conditions, audio formats, and surah coverage. All recordings were standardized to a unified audio format and sampling rate to ensure consistency across the entire dataset. Long-form recordings that contain complete juz’, surahs, or full Qur’an recitations were preserved to enable both verse-level and continuous-speech modeling tasks.

\subsection{Metadata Extraction via Large Language Models}

Due to the absence of consistent and structured metadata across source platforms, we employ large language models (LLMs) to infer and normalize essential annotation fields from unstructured textual descriptions and file-level information.

Titles, descriptions, and related textual attributes are provided to the LLM, which extracts standardized metadata such as surah name and reciter identity. When metadata is incomplete or noisy, the model performs semantic inference to produce the most plausible structured representation.

The resulting normalized metadata enables consistent annotation and supports subsequent dataset curation. The metadata extraction stage employs \textbf{Gemini 2.5 Flash}~\cite{gemini2025} as part of a multi-stage LLM pipeline, where the model is prompted with a structured classification instruction to determine whether a given audio file corresponds to a valid Qur'anic surah recitation and to extract normalized metadata fields such as surah name and reciter identity.



\subsection{Verse-Level Alignment Using Whisper and Quran API}

To obtain precise verse-level segmentation and word-level temporal annotations, we employed an automatic speech recognition (ASR)-driven alignment pipeline. All audio recordings were first processed using the Whisper Large v3 model in combination with WhisperX \cite{bain2023whisperx}, which enables accurate word-level timestamp extraction through forced alignment. The resulting transcriptions were subsequently aligned with the canonical Qur’anic text retrieved from the Quran API.\footnote{\url{https://quranapi.pages.dev/}}

This process enables reliable extraction of verse-level audio segments from long-form recitations and produces temporally synchronized pairs of audio and ground-truth verse text. The complete alignment workflow constitutes the \textbf{Ayah Alignment Module (AAM)}.

\begin{figure}[H]
    \centering
    \includegraphics[width=0.85\textwidth]{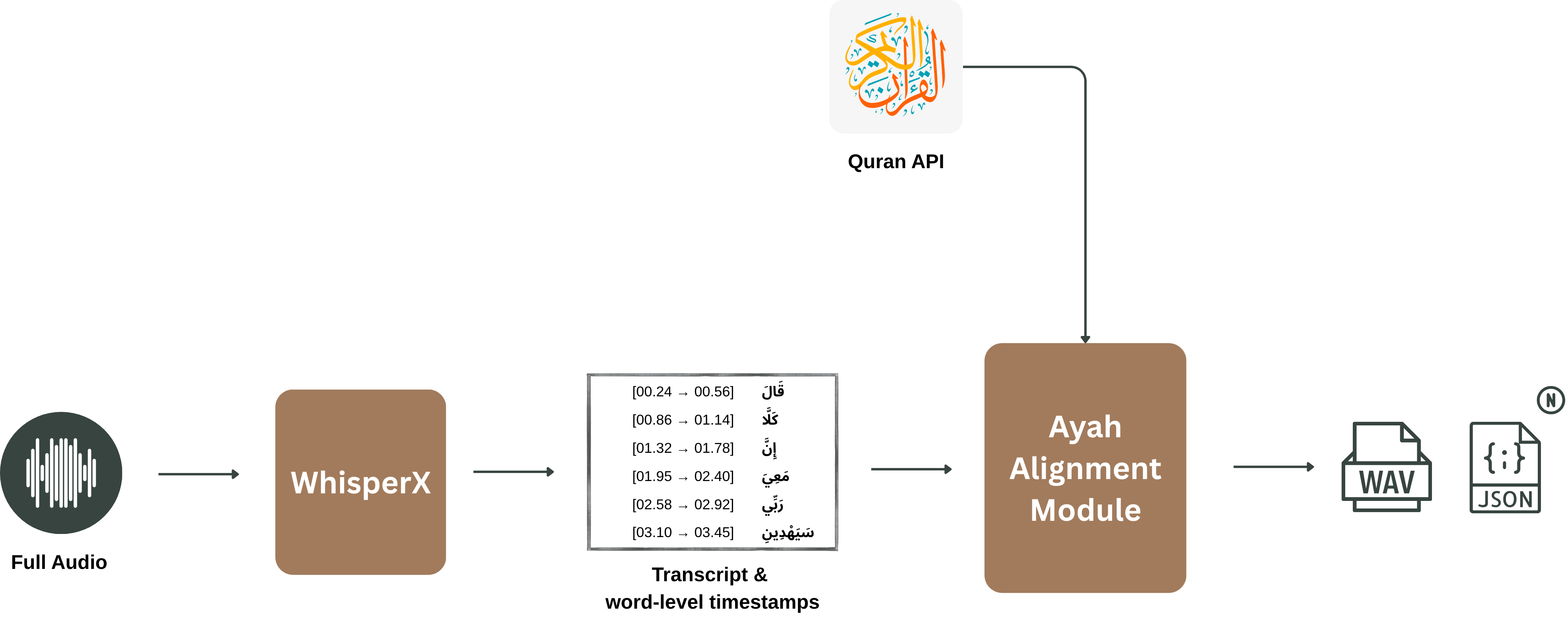}
    \caption{Overview of the verse-level alignment pipeline, illustrating the processing flow from audio input through WhisperX transcription to the Ayah Alignment Module, which generates synchronized verse-level audio segments and structured outputs.}
    \label{fig:alignment_pipeline}
\end{figure}

Within the \textbf{Ayah Alignment Module}, each verse (ayah) of a given surah is iteratively matched against the WhisperX transcription output using a semantic similarity-based approach. Specifically, verse text embeddings are generated using the SILMA AI embedding model \cite{silma2024embedding} and compared to embeddings derived from the corresponding transcription segments.

Cosine similarity is computed between each verse embedding and candidate transcription segments. A verse is considered aligned when the similarity score exceeds a predefined threshold. Upon successful matching, the corresponding start and end timestamps are extracted from the WhisperX output, enabling precise temporal segmentation of the recitation at the verse level.

\begin{figure}[H]
    \centering
    \includegraphics[width=0.85\textwidth]{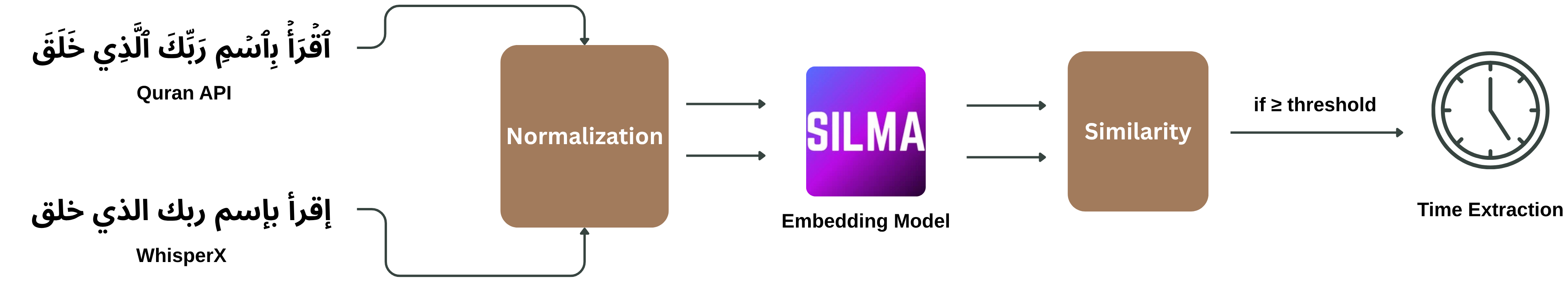}
    \caption{Detailed architecture of the Ayah Alignment Module, illustrating the embedding-based similarity matching and timestamp extraction process.}
    \label{fig:ayah_alignment_module}
\end{figure}

Finally, to maximize alignment precision and ensure that each cropped segment contains only the intended ayah, we incorporated a recitation stop segmentation stage. Following the approach proposed in \cite{ibrahim2025automatic}, a dedicated recitation-boundary detection model was applied to each preliminary cropped segment.

To account for potential boundary underestimation, a temporal buffer of 5 seconds (empirically determined in our experiments) was appended to the end of each cropped audio segment prior to segmentation inference. The segmentation model was then used to detect the precise stopping point of the reciter.

The detected recitation boundary was subsequently reconciled with the corresponding WhisperX timestamp output to refine the segment endpoint. This correction step ensures that the final audio segment terminates exactly at the natural stopping position of the reciter after the verse, eliminating residual speech from subsequent verses or unintended truncation.

This refinement process produces a temporally accurate and semantically clean ayah-level audio segment, forming the final curated output of the alignment pipeline.

\begin{figure}[H]
    \centering
    \includegraphics[width=0.85\textwidth]{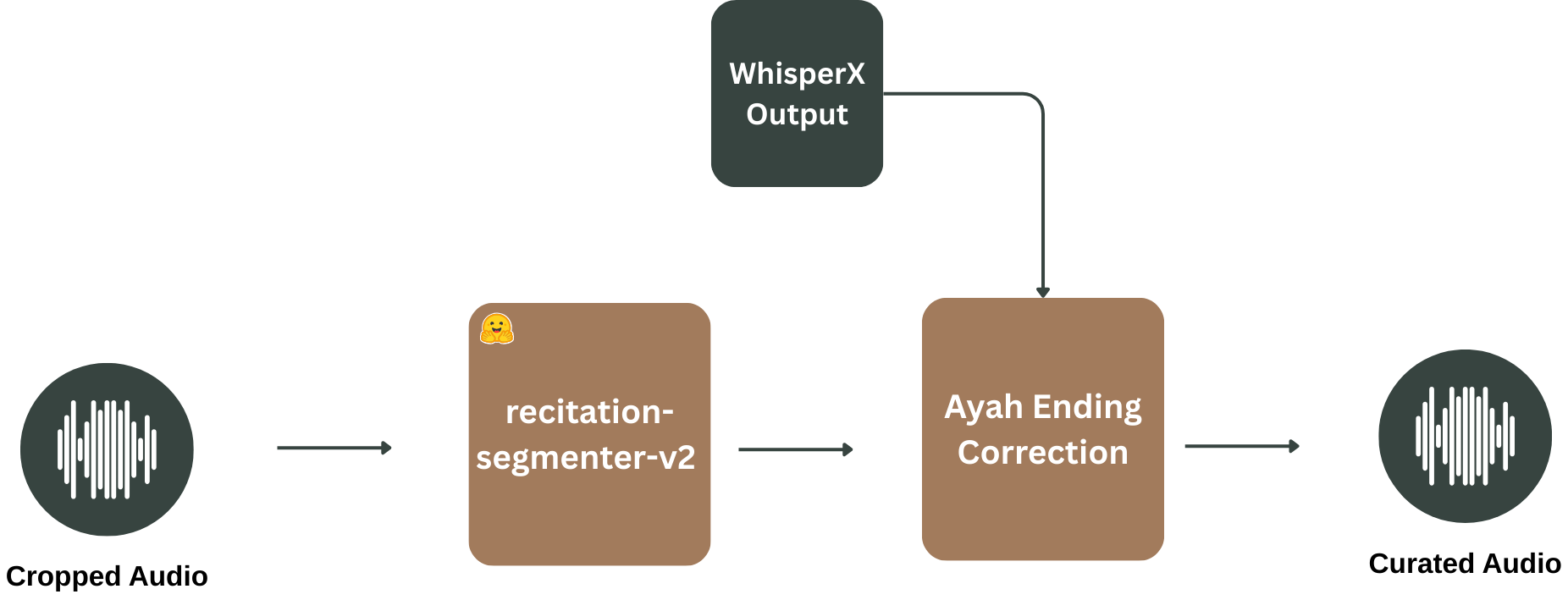}
    \caption{Recitation stop segmentation and boundary correction module for precise ayah-level audio extraction through segmentation inference and timestamp refinement.}
    \label{fig:audio_correction_module}
\end{figure}

\subsection{Dataset Curation}

The final stage of the pipeline focuses on dataset curation to ensure validity and consistency. We apply three complementary mechanisms:

\begin{itemize}
    \item \textbf{LLM-based metadata curation:} Semantic validation using structured metadata.
    \item \textbf{ASR-based curation:} Content verification through verse alignment with canonical Qur’anic text.
    \item \textbf{Deduplication:} Removal of duplicate or near-duplicate recordings.
\end{itemize}

\paragraph{LLM-Based Curation from Metadata}

In the first approach, we leverage a Large Language Model (LLM) as a semantic data curator. The metadata collected during the initial data acquisition stage (e.g., title, description, tags, and other textual attributes) is provided as input to the LLM. The model is then tasked with classifying whether the corresponding audio sample represents a valid Qur’anic surah recitation or not.

This step enables automated semantic validation based solely on contextual metadata, allowing us to filter out mislabeled, incomplete, or non-relevant recordings prior to downstream processing.

\begin{figure}[H]
    \centering
    \includegraphics[width=0.5\textwidth]{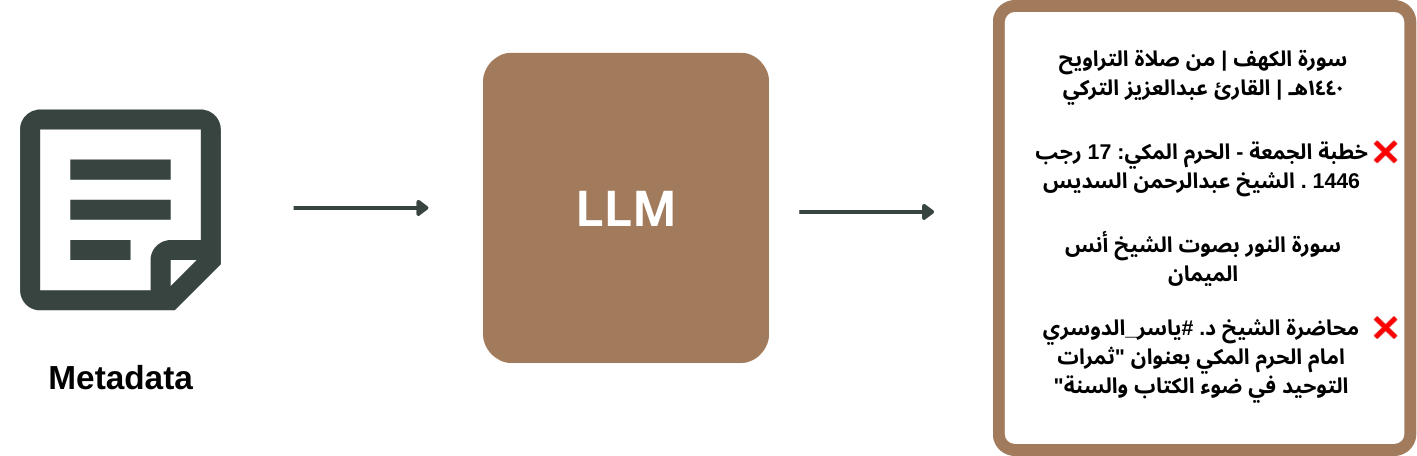}
    \caption{LLM-based metadata classification pipeline for identifying valid surah recitations.}
    \label{fig:data_curation_llm}
\end{figure}

\paragraph{ASR-Based Curation}

Our previously developed ASR-based verse alignment pipeline inherently performs content-level curation. Since alignment is conducted against canonical verses obtained from the Qur’an API, only recordings containing valid surah recitation can be successfully aligned. 

As a result, non-recitation content (e.g., sermons or other speech segments) naturally fails the alignment process and is excluded. Thus, verse-level alignment implicitly serves as an additional data curation mechanism.

\paragraph{Deduplication}

Given the large-scale data collection process, duplicate recordings—particularly for the same reciter and verse (ayah)—are likely to occur. To ensure dataset integrity, we perform a dedicated deduplication step.

We first group samples that share the same reciter and verse. For each group, we extract audio embeddings using the Efficient Audio Transformer (EAT) \cite{chen2024eat}. We then compute pairwise cosine similarity between recordings within the same group. If the similarity exceeds a predefined threshold (0.9 in our experiments), the recordings are considered duplicates and merged accordingly.

Inspired by \cite{oquab2023dinov2}, we model duplicate relationships as a graph and extract connected components using a scalable disjoint-set (union–find) data structure. Each connected component corresponds to a duplicate group, from which we retain a single representative recording and discard the remaining samples.

\begin{figure}[H]
    \centering
    \includegraphics[width=0.8\textwidth]{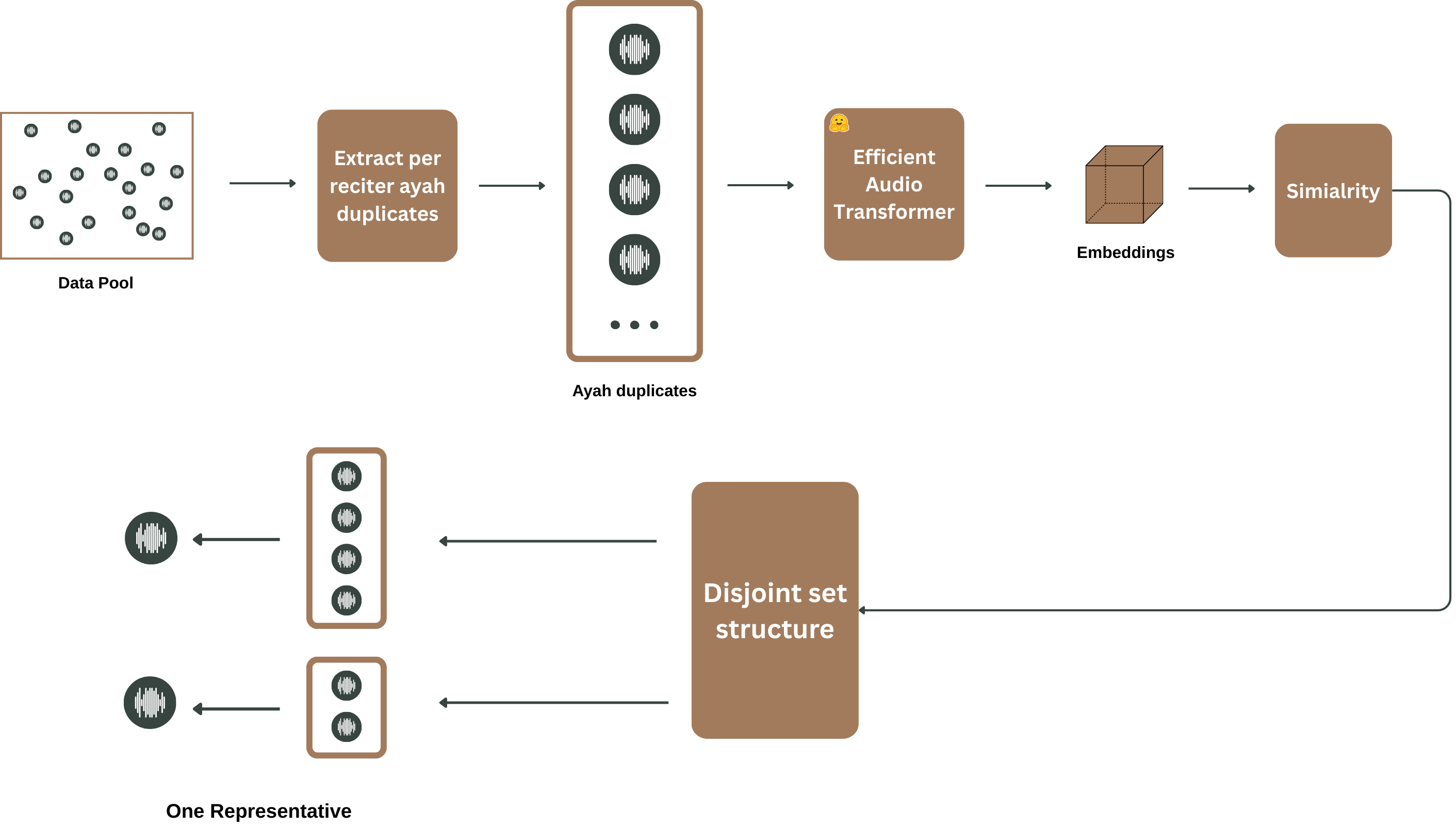}
    \caption{Deduplication pipeline using EAT embeddings, cosine similarity, and connected component extraction.}
    \label{fig:data_curation_dedup}
\end{figure}

\section{Pipeline Quality Evaluation}

A core component of the Tadabur construction pipeline is the Ayah Alignment Module (AAM), which is responsible for identifying and segmenting individual verse-level audio from long-form recitations. The quality of the resulting dataset is directly dependent on the reliability of this alignment stage. In this section, we evaluate two critical design dimensions of the AAM: (1) the \textbf{alignment method} used to match transcription outputs to canonical Qur'anic text, and (2) the \textbf{ASR model} used to generate transcriptions.

\subsection{Evaluation Setup}
We evaluate alignment coverage across five well-known reciters: Abd al-Basit Abd al-Samad, Abdulmohsen Al-Qasim, Abd al-Rahman al-Sudais, Saud Al-Shuraim, and Yasser Al-Dosari. Each reciter has a ground truth corresponding to complete Qur'an coverage across all surahs and ayahs. Importantly, the audio recordings used in this evaluation were \textbf{not seen during fine-tuning} of the Tadabur model, ensuring an unbiased assessment of the alignment pipeline on unseen data. We define \textbf{alignment coverage} as the percentage of ayahs successfully identified and segmented by the pipeline:

\[
\text{Coverage} = \frac{\text{Aligned Ayahs}}{\text{Total Ayahs}} \times 100
\]

The evaluation set for each reciter was curated specifically for this purpose: recordings were gathered independently, deduplicated, and cleaned to ensure that each ayah appears \textbf{exactly once}. Coverage, therefore, reflects the pipeline's ability to correctly identify and segment unique ayahs, with no inflation from repeated recordings.

We evaluate two alignment methods and three ASR models:

\begin{enumerate}[label=\textbf{\arabic*.}]
    \item \textbf{Alignment methods:}
    \begin{enumerate}[label=(\alph*)]
        \item \textit{SILMA Embedding} — semantic similarity via the SILMA AI embedding model \cite{silma2024embedding}
        \item \textit{Fuzzy Matching} — text-based string similarity baseline
    \end{enumerate}

    \item \textbf{ASR models:}
    \begin{enumerate}[label=(\alph*)]
        \item \textit{Tadabur fine-tuned model} (ours)
        \item \textit{Whisper-Quran}~\cite{tarteel2022whisper} — an existing Whisper model fine-tuned on Qur'anic data
        \item \textit{Whisper Small} — standard base model without domain adaptation
    \end{enumerate}
\end{enumerate}

\subsection{Results}

Table~\ref{tab:pipeline_quality} reports alignment coverage for each configuration. We additionally include the average coverage across all five reciters for a concise summary.

\begin{table}[H]
\centering
\small
\setlength{\tabcolsep}{5pt}
\renewcommand{\arraystretch}{1.15}
\begin{tabular}{llccccc|c}
\toprule
\textbf{Alignment} & \textbf{ASR Model}
  & \rotatebox{60}{\textbf{Abd al-Basit}}
  & \rotatebox{60}{\textbf{Al-Qasim}}
  & \rotatebox{60}{\textbf{Al-Sudais}}
  & \rotatebox{60}{\textbf{Al-Shuraim}}
  & \rotatebox{60}{\textbf{Al-Dosari}}
  & \textbf{Avg.} \\
\midrule
SILMA Emb.  & Tadabur (Ours)   & 95.61 & 97.73 & 94.47 & 98.14 & 97.18 & \textbf{96.63} \\
SILMA Emb.  & Whisper-Quran    & 97.25 & 96.70 & 92.69 & 92.61 & 98.23 & 95.50 \\
SILMA Emb.  & Whisper Small    & 79.67 & 82.35 & 76.34 & 86.95 & 87.56 & 82.57 \\
\midrule
Fuzzy Match & Tadabur (Ours)   & 80.76 & 89.57 & 83.48 & 92.00 & 84.35 & 86.03 \\
Fuzzy Match & Whisper-Quran    & 87.29 & 91.42 & 83.33 & 84.41 & 89.68 & 87.23 \\
Fuzzy Match & Whisper Small    & 69.75 & 73.26 & 67.00 & 78.25 & 75.75 & 72.80 \\
\bottomrule
\end{tabular}
\vspace{6pt}
\caption{Ayah alignment coverage (\%) across five reciters under different alignment methods and ASR models.}
\label{tab:pipeline_quality}
\end{table}

\subsection{Analysis}

\paragraph{Alignment Method.}
SILMA Embedding-based alignment consistently outperforms fuzzy text matching across all reciters and ASR models. With the Tadabur fine-tuned model, SILMA achieves an average coverage of \textbf{96.63\%}, compared to \textbf{86.03\%} for fuzzy matching — a gap of more than \textbf{10 percentage points}. This advantage holds across all three ASR backbones, confirming that semantic-level matching is significantly more robust to the phonological variations, elongated phoneme durations, and recitation-specific disfluencies that characterize Qur'anic speech, which cause brittle surface-form text matching to fail.

\paragraph{ASR Model.}
The choice of ASR model has a clear and consistent impact on alignment coverage. Whisper Small, without any domain adaptation, achieves the lowest coverage across both alignment methods: \textbf{82.57\%} under SILMA Embedding and \textbf{72.80\%} under Fuzzy Matching. This degradation reflects the model's limited exposure to Qur'anic Arabic during pre-training, resulting in transcription errors that propagate into the alignment stage.

Among domain-adapted models, the Tadabur fine-tuned model achieves the highest average coverage of \textbf{96.63\%} under SILMA Embedding, marginally outperforming Whisper-Quran~\cite{tarteel2022whisper} (\textbf{95.50\%}). Notably, under Fuzzy Matching, Whisper-Quran (\textbf{87.23\%}) slightly exceeds Tadabur (\textbf{86.03\%}), suggesting that both models are competitive and that the choice of alignment method plays a larger role than the specific ASR backbone when both are domain-adapted.

\paragraph{Summary.}
These results validate the two core design choices of the AAM: the use of semantic embedding-based alignment over surface-form text matching, and the use of a domain-adapted Qur'anic ASR backbone. The adopted configuration — SILMA Embedding with the Tadabur fine-tuned model — achieves the highest alignment coverage and serves as the basis for the full Tadabur dataset construction pipeline.

\section{Dataset Statistics}
The \textbf{Tadabur} dataset provides one of the most extensive and diverse collections of Quranic recitation audio available to date. This section presents a quantitative overview of the dataset's main characteristics. Note that the distribution of recordings is not uniform across reciters, as the dataset reflects the natural variability in the availability of audio sources during collection.

\subsection{Dataset Size}

The final dataset contains:

\begin{itemize}
\item More than \textbf{1400+ hours} of verse-level annotated audio
\item Over \textbf{600} distinct reciters spanning a wide range of ages, dialects, and recitation traditions
\item Automatically generated word-level temporal alignments and structured metadata
\end{itemize}

Compared to previously available Quranic datasets, \textbf{Tadabur} offers a substantially larger scale in both total audio duration and reciter diversity. Table~\ref{tab:dataset_comparison} provides a quantitative comparison between Tadabur and the most widely used publicly available Quranic datasets.

\begin{table}[H]
\centering
\small
\setlength{\tabcolsep}{6pt}
\begin{tabular}{lcccc}
\toprule
\textbf{Dataset Name} & \textbf{Segments} & \textbf{Reciters} & \textbf{Transcription} & \textbf{Word-Level Align.} \\
\midrule
Quran Recitations for Audio Classification 
& 6,689 & 12 & $\times$ & $\times$ \\

Quran Speech-to-Text (SLR132) 
& 226,129 & 30 & $\checkmark$ & $\times$ \\

Buraaq Quran Audio--Text Dataset 
& 187,080 & 30 & $\checkmark$ & $\times$ \\

\textbf{Tadabur (Ours)} 
& \textbf{365,000+} & \textbf{600+} & $\checkmark$ & $\checkmark$ \\
\bottomrule
\end{tabular}

\vspace{6pt}
\caption{Comparison between Tadabur and existing publicly available Quran datasets.}
\label{tab:dataset_comparison}
\end{table}

\subsection{Reciter Diversity}

Reciter diversity is one of the defining strengths of the \textbf{Tadabur} dataset. In addition to covering a wide range of dialects and recitation traditions, the dataset includes multiple recordings of the \textit{same} surah and ayah for many reciters. These natural variations arise from differences in recording sessions, recitation pace, melodic choices, and acoustic environments. As a result, Tadabur captures not only cross-reciter diversity but also within-reciter stylistic variation, providing a richer representation of real-world Quranic recitation. This diversity enables research on robustness to stylistic shifts, modeling recitation styles, multi-take variability, and speaker-specific prosodic patterns.

\begin{figure}[H]
\centering
\small
\begin{tikzpicture}
\begin{axis}[
    ybar,
    bar width=18pt,
    width=0.8\linewidth,
    height=6cm,
    ylabel={Number of Reciters},
    symbolic x coords={Kaggle, SLR132, Buraaq, Tadabur (Ours)},
    xtick=data,
    axis lines=box,        
    ymajorgrids=false,
    nodes near coords,
    nodes near coords style={font=\small},
    enlarge x limits=0.15
]

\addplot[
    fill={rgb,255:red,55;green,68;blue,64},
    draw=black
] coordinates {
    (Kaggle,12)
    (SLR132,30)
    (Buraaq,30)
    (Tadabur (Ours),600)
};

\end{axis}
\end{tikzpicture}

\vspace{4pt}
\caption{Number of reciters across major publicly available Quran datasets.}
\label{fig:reciter_comparison}
\end{figure}
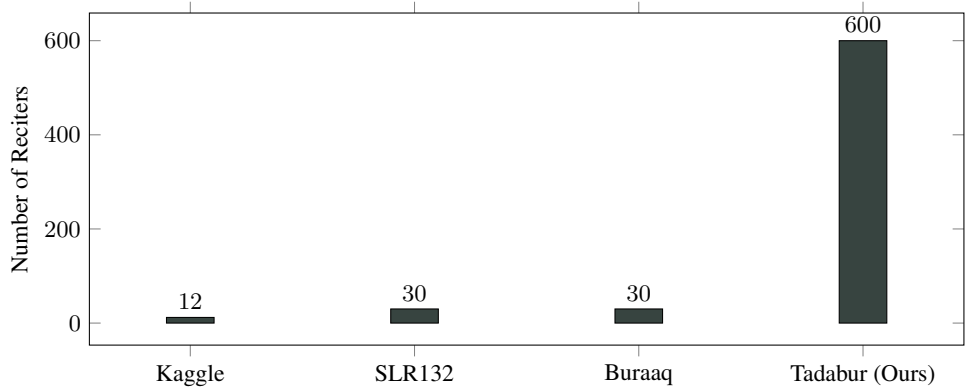

Figure~\ref{fig:reciter_comparison} illustrates how Tadabur compares to existing Quranic datasets in terms of the number of unique reciters.

\section{Models Evaluation Against Tadabur}

To demonstrate the utility of Tadabur as a benchmark for Qur'anic ASR, we evaluate a diverse set of publicly available speech recognition models on the full dataset. The models span a range of architectures, parameter scales, and levels of Arabic or Qur'anic domain adaptation, providing a comprehensive baseline for future research.

\subsection{Models}

We evaluate the following eight models:

\begin{itemize}
    \item \textbf{Whisper-Quran}~\cite{tarteel2022whisper} — a Whisper Base model fine-tuned specifically on Qur'anic Arabic speech (74M parameters).
    \item \textbf{Whisper Small}~\cite{radford2023whisper} — the standard OpenAI Whisper Small model without domain adaptation (244M parameters).
    \item \textbf{Wav2Vec2 XLSR-53 Arabic}~\cite{grosman2021xlsr53-large-arabic} — the XLSR-53 multilingual model fine-tuned on Arabic speech (300M parameters).
    \item \textbf{MMS 1B}~\cite{pratap2023mms} — Meta's Massively Multilingual Speech model, covering 1,000+ languages including Arabic (1B parameters).
    \item \textbf{Qwen3-ASR-1.7B}~\cite{qwen3asr2026} — Alibaba's state-of-the-art multilingual ASR model supporting 52 languages (1.7B parameters).
    \item \textbf{Cohere Transcribe}~\cite{coheretranscribe2026} — Cohere Labs' dedicated ASR model achieving state-of-the-art on the Open ASR Leaderboard (2B parameters).
    \item \textbf{Voxtral Mini}~\cite{voxtral2026} — Mistral AI's natively streaming multilingual ASR model supporting 13 languages (4B parameters).
    \item \textbf{VibeVoice-ASR}~\cite{vibevoice2026} — Microsoft Research's unified speech-to-text model supporting 50+ languages with long-form audio processing (7B parameters).
\end{itemize}

\subsection{Evaluation Metric}
We evaluate all models using \textbf{Word Error Rate (WER)} and \textbf{Character Error Rate (CER)}, which are the standard metrics for ASR evaluation. Prior to computing both metrics, all predictions and labels were normalized by removing diacritics (tashkeel), Qur'anic punctuation marks (waqf signs), and orthographic variants specific to the Uthmani script, ensuring that error rates reflect lexical recognition accuracy rather than orthographic convention differences. Both are computed against the canonical Qur'anic text provided in each sample's metadata. Lower values indicate better transcription accuracy.

\subsection{Results}

Table~\ref{tab:asr_benchmark} reports WER and CER for all evaluated models on Tadabur, sorted by WER.

\begin{table}[H]
\centering
\small
\setlength{\tabcolsep}{8pt}
\renewcommand{\arraystretch}{1.15}
\begin{tabular}{lrcc}
\toprule
\textbf{Model} & \textbf{Size} & \textbf{WER (\%) $\downarrow$} & \textbf{CER (\%) $\downarrow$} \\
\midrule
Whisper-Quran~\cite{tarteel2022whisper}           & 74M   & \textbf{8.7}  & \textbf{6.5}  \\
Cohere Transcribe~\cite{coheretranscribe2026}      & 2B    & 11.2          & 8.1           \\
Voxtral Mini~\cite{voxtral2026}                    & 4B    & 15.1          & 11.2          \\
VibeVoice-ASR~\cite{vibevoice2026}                 & 7B    & 24.3          & 14.0          \\
Qwen3-ASR-1.7B~\cite{qwen3asr2026}                & 1.7B  & 25.2          & 9.9           \\
Whisper Small~\cite{radford2023whisper}            & 244M  & 29.2          & 16.1          \\
MMS 1B~\cite{pratap2023mms}                        & 1B    & 51.1          & 16.6          \\
Wav2Vec2 XLSR-53 Arabic~\cite{grosman2021xlsr53-large-arabic}    & 300M  & 57.4          & 21.9          \\
\bottomrule
\end{tabular}
\vspace{6pt}
\caption{ASR evaluation results on Tadabur. Models are sorted by WER. WER and CER are computed against the canonical Qur'anic text after normalization.}
\label{tab:asr_benchmark}
\end{table}

\subsection{Analysis}

The results reveal a clear pattern: \textbf{domain adaptation outweighs model size} in Qur'anic ASR. Whisper-Quran, despite being the smallest model at 74M parameters, achieves the best WER of \textbf{8.7\%} and CER of \textbf{6.5\%}, substantially outperforming much larger general-purpose models. Among those, Cohere Transcribe (\textbf{11.2\%}) and Voxtral Mini (\textbf{15.1\%}) perform most competitively — notably, Cohere Transcribe was not trained on any Qur'anic data, yet achieves strong results, reflecting the strength of its large-scale multilingual pre-training. In contrast, MMS 1B (\textbf{51.1\%}) and Wav2Vec2 XLSR-53 Arabic (\textbf{57.4\%}) perform poorly, confirming that neither multilingual nor Arabic-specific training generalizes reliably to the phonologically distinct domain of Qur'anic recitation.

\newpage

\section{Licensing and Ethical Considerations}
\textbf{Tadabur} is published as an open-source dataset to support research in Arabic audio and speech technologies. The dataset is freely accessible to the research community and is intended to lower the barrier to working with large-scale, well-annotated Qur'anic speech data. Alongside the audio and annotations, we release metadata files to make it easier to explore, analyze, and build on the dataset.

Given the central religious significance of the Qur'an, we emphasize that Tadabur is intended for respectful and beneficial use, particularly in educational, accessibility, and academic research contexts. Users are expected to avoid applications that would constitute mocking, distortion, or otherwise disrespectful manipulation of Qur'anic recitation, and to adhere to relevant cultural and religious norms in their work.

\section{Limitations}
Although Tadabur is the largest Quranic audio dataset to date, it has some limitations that future work can address.

The first limitation is that some reciters do not have audio recordings for every ayah. This is either because a reciter had a small number of recordings available during data collection or because the processing pipeline failed to correctly match the audio to the right ayah, which was mostly caused by errors in the speech recognition step.

The second limitation is that the word-level timestamps are not always precise. This is because the alignment model used was not originally built for Quranic audio and therefore struggles with the unique pronunciation and recitation style found in the Quran.

\bibliographystyle{IEEEtran}   
\bibliography{references}

\end{document}